\documentclass[aps,twocolumn,superscriptaddress,showpacs,floatfix]{revtex4}
\usepackage[dvips]{graphicx}
\usepackage{amsmath}

\begin{document}

\title{Superfluid Interfaces in Quantum Solids}

\author{Evgeni Burovski}
\author{Evgeni Kozik}
\affiliation{Department of Physics, University of Massachusetts,
Amherst, MA 01003}
\author{Anatoly Kuklov}
\affiliation{Department of Engineering Science and Physics, The College %
of Staten Island, City University of New York, Staten Island, NY
10314}
\author{Nikolay Prokof'ev}
\author{Boris Svistunov}
\affiliation{Department of Physics, University of Massachusetts,
Amherst, MA 01003}%
\affiliation{Russian Research Center
``Kurchatov Institute'', 123182 Moscow, Russia}

\begin{abstract}
One scenario for the non-classical moment of inertia of solid
$^4$He discovered by Kim and Chan [Nature, {\bf 427}, 225 (2004)]
is the superfluidity of micro-crystallite interfaces. On the basis
of the most simple model of a quantum crystal---the checkerboard
lattice solid---we show that the superfluidity of interfaces
between solid domains can exist in a wide range of parameters. At
strong enough inter-particle interaction, a superfluid interface
becomes an insulator via a quantum phase transition. Under the
conditions of particle-hole symmetry, the transition is of the
standard $U(1)$ universality class in 3D, while in 2D the onset of
superfluidity is accompanied by the interface roughening, driven
by fractionally charged topological excitations.
\end{abstract}


\maketitle

Recent observation by Kim and Chan \cite{KimChan04} of
non-classical moment of inertia (NCMI) of $^4$He at pressures
significantly higher than the solidification point is a
breathtaking result, especially striking in view of the
theorem-like theoretical arguments against existence of
commensurate supersolids \cite{PS-SSOLID}. The fact of
commensurability---equivalently, one may put it as the fact of
absence of vacancies, or interstitials, or both---of the
equilibrium solid $^4$He at $T=0$ is supported by an extensive
experimental work over the past several decades (for review, see,
e.g., \cite{meisel}), as well as by the most recent experimental
and numeric studies \cite{Beamish04,Ceperley}.  The
commensurability of solid $^4$He rules out NCMI based on
Bose-Einstein condensation of vacancies
\cite{AndrLifsh69-Chester70}. Two of us have proposed recently
\cite{PS-SSOLID} that NCMI might be due to the superfluidity of
interfaces between $^4$He crystallites. At present, the weak point
of this hypothesis is the absence of a theoretical analysis and/or
direct experimental evidence of the superfluidity in the walls
separating insulating domains.

The problem of interface superfluidity in a quantum solid is of
significant general interest on its own, being potentially
relevant not only to the solid $^4$He polycrystal, but also to the
properties of domain walls in spin arrays and ultracold atoms in
optical lattices.


In this Letter, we present a proof-of-principle study of
superfluidity in interfaces between insulating domains with broken
translation symmetry (solids). We address the problem by studying
the checkerboard lattice solid (CB). We start with giving a simple
illustrative theoretical argument that at least under certain
limiting conditions the domain wall in our system {\it has} to be
superfluid. Our numeric simulations of 2D and 3D models reveal
superfluidity of the CB domain walls in a large range of
parameters. We pay special attention to the study of the
superfluid (SF)--insulator (I) quantum phase transition in the
interface. In 3D, the transition turns out to be in the $U(1)$
universality class. In 2D, we conclude that the I-to-SF transition
in the wall is driven by proliferation of topological excitations
that carry the fractional particle charge $1/2$, as well as the
quantum of the interface shift in the transversal direction. The
latter circumstance results in an interesting effect: The
transition is accompanied by the wall {\it roughening}.

The simplest system featuring both superfluid and CB phases is
that of the hard-core lattice bosons with nearest-neighbor
repulsion, at half-integer filling factor (see, e.g.,
\cite{Troyer} and references therein). The model can be exactly
mapped onto spin-1/2 XXZ antiferromagnet, which leads to the
following correspondence. The CB phase is equivalent to the
easy-axis antiferromagnet (characterized by the broken $Z_2$
symmetry), while the SF phase is identified with the easy-plane
antiferromagnet (characterized by broken $U(1)$ symmetry).
Correspondingly, in terms of the N\'eel vector, $\vec{S}$, the CB
order parameter is $M=S_z$, while the SF order parameter is
$\Psi=S_x+i S_y$. Generically, the groundstate of the model is
either SF or CB, depending on the Hamiltonian parameters. There is
also a special $SU(2)$-symmetric Heisenberg point. In 2D and 3D
groundstates of the Heisenberg Hamiltonian the $SU(2)$ symmetry is
broken, so that the vector $\vec{S}$ is non-zero and can point at
any direction.

Let us take now an easy-axis Hamiltonian that is very close to the
Heisenberg point ($T=0$), and create two large domains, $S_z=M$
and $S_z=-M$. What is the structure of the domain wall? Being
close to the Heisenberg point, we are forced to conclude that the
wall is very thick in the transverse direction, with the vector
$\vec{S}$ well defied locally inside the wall and evolving
smoothly from $(0,0,M)$ to $(0,0,-M)$ across the wall. The
energetic cost is controlled by the closeness to the Heisenberg
point and can be rendered arbitrarily small. In the middle of the
wall, $\vec{S}=(S_x,S_y,0)$, which means that the wall is
characterized by broken $U(1)$ symmetry implying superfluidity in
the bosonic case. As the system is taken deeper into the solid
state in the bulk, it becomes energetically favorable to suppress
the module of $\vec{S}$ to zero in the wall, which means an
insulating state of the wall. Similar transformation of the domain
wall---from the Bloch-type, where $|\vec{S}|$ is finite
everywhere, to the Ising-type, where $|\vec{S}|=0$ in the middle
of the wall---takes place in classical ferromagnets close to the
Curie point \cite{Ginzburg}.

\begin{figure*}[tbh]
\includegraphics[width=0.95\textwidth, keepaspectratio=true]{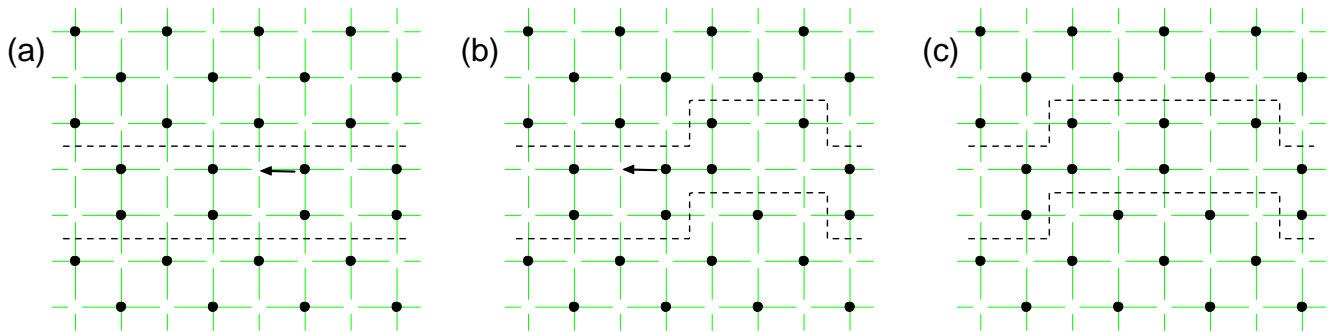}
\caption{A sketch of the wall roughening mechanism due to spinons.
(a): a ``frozen'' wall (the arrow shows a single-particle hopping
event, which generates the configuration (b), and the dashed lines
are to facilitate the wall gazing); (b): a configuration,
featuring a pair of spinons; (c): with a single-particle hopping
over one lattice site the spinon is shifted by two lattice
periods, which means that the spinon particle charge is $1/2$.}
\label{fig:spinons}
\end{figure*}

Once the superfluidity of a domain wall is established in the
limiting case, one may expect that it can take place under more
general conditions, especially in view of the following, almost
obvious, energetic argument. Due to geometrical frustration, the
energy cost to translate a particle along the wall is less than in
the bulk. For example, in the case of the 2D CB solid, the
particle jump increases the energy by $\sim 3V$, where $V>0$ is
the interaction energy between two close neighbors. A particle at
the wall [see, e.g., Fig.~\ref{fig:spinons}(a)] already has
another particle as its neighbor. After jumping one step along the
wall, it acquires only two close neighbors, and the energy
increase due to such a jump is only $\sim V$.

A point of concern, however, is that generically the transition
from SF to CB is of the first order so that for a given system or
range of parameters it may turn out that the interface between two
insulating domains is {\it always} in the insulating state.

To get an idea of how likely it is to get the interface between
the two CB domains superfluid, we simulate a domain wall in the
bond-current model---a discrete-imaginary-time analog of a
$(d+1)$-dimensional worldline representation of a quantum bosonic
or spin system in $d$ spatial dimensions \cite{Wallin94}. The
Hamiltonian of the model reads
\begin{equation}
H =
t \sum_{\mathbf{n}} \sum_{\alpha=1}^d J^2_{\mathbf{n},\alpha} +
\frac{p}{2} \sum_{\langle \mathbf{n},\mathbf{m} \rangle } J_{\mathbf{n}, \tau}
J_{\mathbf{m},\tau}\; . \label{j-curr}
\end{equation}
Here the integer vector $\mathbf{n}=(n_1, \ldots ,n_d,n_\tau)$
labels sites of the $(d+1)$-dimensional cubic lattice,
$\alpha=1,\ldots,d$ enumerates the spatial directions, and $\tau$
denotes the temporal direction. Currents $J_{\mathbf{n},\alpha}$
and $J_{\mathbf{n},\tau}$ are integers associated with bonds
adjacent to the site $\mathbf{n}$ in directions $\alpha$ and
$\tau$, respectively. The summation in the second term runs over
all pairs of temporal bonds having a common plaquette. The
configurations of bond currents are subject to the zero-divergence
constraint. Without loss of universality, we restrict the values
of bond currents to $J_{\mathbf{n},\alpha}=0, \pm 1$ and
$J_{\mathbf{n},\tau}= \pm 1$. The zero-divergence constraint then
reads
\begin{equation}
\sum_{\alpha} \left( J_{\mathbf{n},\alpha} +
J_{\mathbf{n},-\alpha} \right) +
\frac{1}{2} \left( J_{\mathbf{n},\tau} +
J_{\mathbf{n},-\tau} \right)%
= 0 \; ,
 \label{zero-div}
\end{equation}
where the negative sign means the opposite direction, so that
$J_{\mathbf{n},-\alpha} \equiv -
J_{(\mathbf{n}-\hat{\alpha}),\alpha}$ and
$J_{\mathbf{n},-\tau}\equiv - J_{(\mathbf{n}-\hat{\tau}),\tau}$;
the hats stand for unit vectors in the corresponding directions.

We simulate the model (\ref{j-curr}), using Worm algorithm
\cite{worm}, in the range of parameters where the bulk is deep
into the solid regime. To automatically create a domain wall, we
take a lattice with periodic boundary conditions and an odd number
of sites in the spatial direction $x$ (system sizes in all the
other directions are even). The domain wall superfluidity
manifests itself as a non-zero mean square of winding numbers
\cite{Ceperley87} in the direction(s) parallel to the wall:
$\langle W^2_\| \rangle \neq 0$, while $\langle W^2_x \rangle =
0$. More specifically, for a 3D system with the domain wall in the
$yz$ plane, the superfluid stiffness is given by $ \rho_s =
\langle W^2_{y} + W^2_{z} \rangle / 2L_{\tau}, $ and likewise for
the compressibility. In a 2D system with the domain wall in the
$y$ direction the statistics of winding numbers is essentially
discrete,$P(W_y) \propto \exp\left[ - (L_y/L_\tau) W^2_y/ 2\rho_s
\right]$, and an appropriate estimator for the superfluid
stiffness is $\rho_s^{-1} = -2 (L_\tau / L_y) \ln [ P(W_y=1) /
P(W_y=0)]$, and likewise for the compressibility. Here $L_\tau$
and $L_y$ are the linear system sizes in the corresponding
directions. In our simulations, we set $L_y=L_z=L_\tau =L$ and
$L_x =L +1$.

In 3D, we found that for $t \lesssim 1$ the interfaces are never
superfluid. Simulations at $t=1.3$ revealed a first-order SF-CB
transition in the bulk at $p\approx 0.2$ with the interface
remaining superfluid till $p$ becomes equal to $p_c=0.27115(5)$.
In 2D, the simulations were performed at $t=0.9$. The first-order
bulk SF--CB transition was found at $p\approx 0.5$, while the
interface becomes insulating only at $p_c=0.7633(5)$.

A superfluid interface embedded into a $d$-dimensional solid is a
$(d-1)$-dimensional superfluid, and the scenario for the quantum
phase transition from SF to I state in such a system is
interesting on its own. Our model, Eq.~(\ref{j-curr}), has a
particle-hole symmetry. Thus, commensurability should play a key
part in the criticality \cite{rem1}. Fundamentally, there are two
qualitatively different cases, depending on whether the interface
is {\it smooth} or {\it rough}. If the interface is centered at
$x=0$ and $x({\vec \rho})$ is its instantaneous shape (${\vec
\rho}$ is the vector in the hyperplane perpendicular to the $x$
axis), then, by definition, $\langle \,x({\vec \rho})^2\, \rangle
\sim 1$ for a smooth  interface, while for a rough interface
$\langle \, x({\vec \rho})^2\, \rangle $ is macroscopically large
(scales as some power of the system size). For a smooth interface,
the CB environment plays a role of a periodic external potential
that doubles the interface unit cell. This means that a smooth
interface can be treated as a commensurate system with an {\it
integer} (unity) filling factor. Its SF--I transition then
corresponds to the superfluid--Mott insulator (MI) transition at
integer filling, known to be of the $U(1)$ universality class
\cite{FWGF89}. If the interface is rough, then the effect of the
solid environment is averaged out by the zero-point fluctuations
of $x({\vec \rho})$ and the effective filling factor for the
interface remains {\it half-integer}, with corresponding
implications for the universality class of the SF--I quantum phase
transition.

In our simulations, we observe the smooth-interface scenario in 3D
and the rough-interface scenario in 2D.

\begin{figure}[htb]
\includegraphics[width = 0.85\columnwidth,keepaspectratio=true]{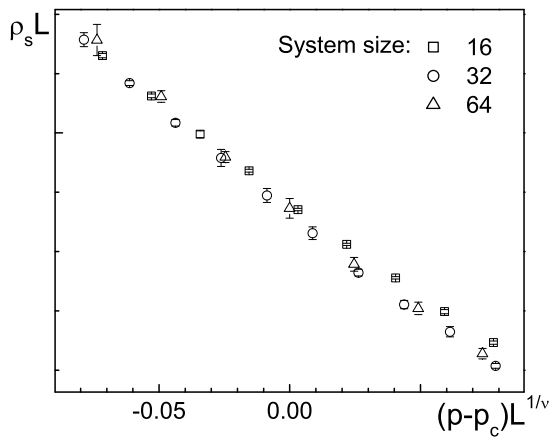}
\caption{Criticality for the interface in a 3D system. The data
for $\rho_s$ as a function of the parameter $p$ is consistent with
the exponent $\nu\approx 0.671$ of the $U(1)$
universality class.}%
\label{fig:Col3d}
\end{figure}

In Fig.~\ref{fig:Col3d} we present the 3D data in the vicinity of
the critical point, $p_c=0.2711(5)$. A very good data collapse
with the critical exponent $\nu \approx 0.671$ of the $U(1)$
universality class is indicative of the standard SF--MI scenario.


In 2D, the interface forms a 1D Luttinger liquid, the main
characteristic of which is the dimensionless parameter $g = \pi
\sqrt{\rho_s \kappa}$, where $\rho_s$ and $\kappa$ are the 1D
superfluid stiffness and compressibility, respectively. In a
Luttinger liquid with a filling factor $1/m$ ($m$ is an integer)
the SF--I (Kosterlitz-Thouless type) transition takes place at
$g=g_c=2/m^2$ \cite{g}. As we argued above, a smooth interface
implies $m=1$ and, correspondingly, criticality at $g_c=2$. The
results of our simulation show that the SF--I transition actually
takes place at $g_c=1/2$, corresponding to the half-filling case,
implying the rough interface scenario.

\begin{figure}[htb]
\includegraphics[width=0.85\columnwidth,keepaspectratio=true]{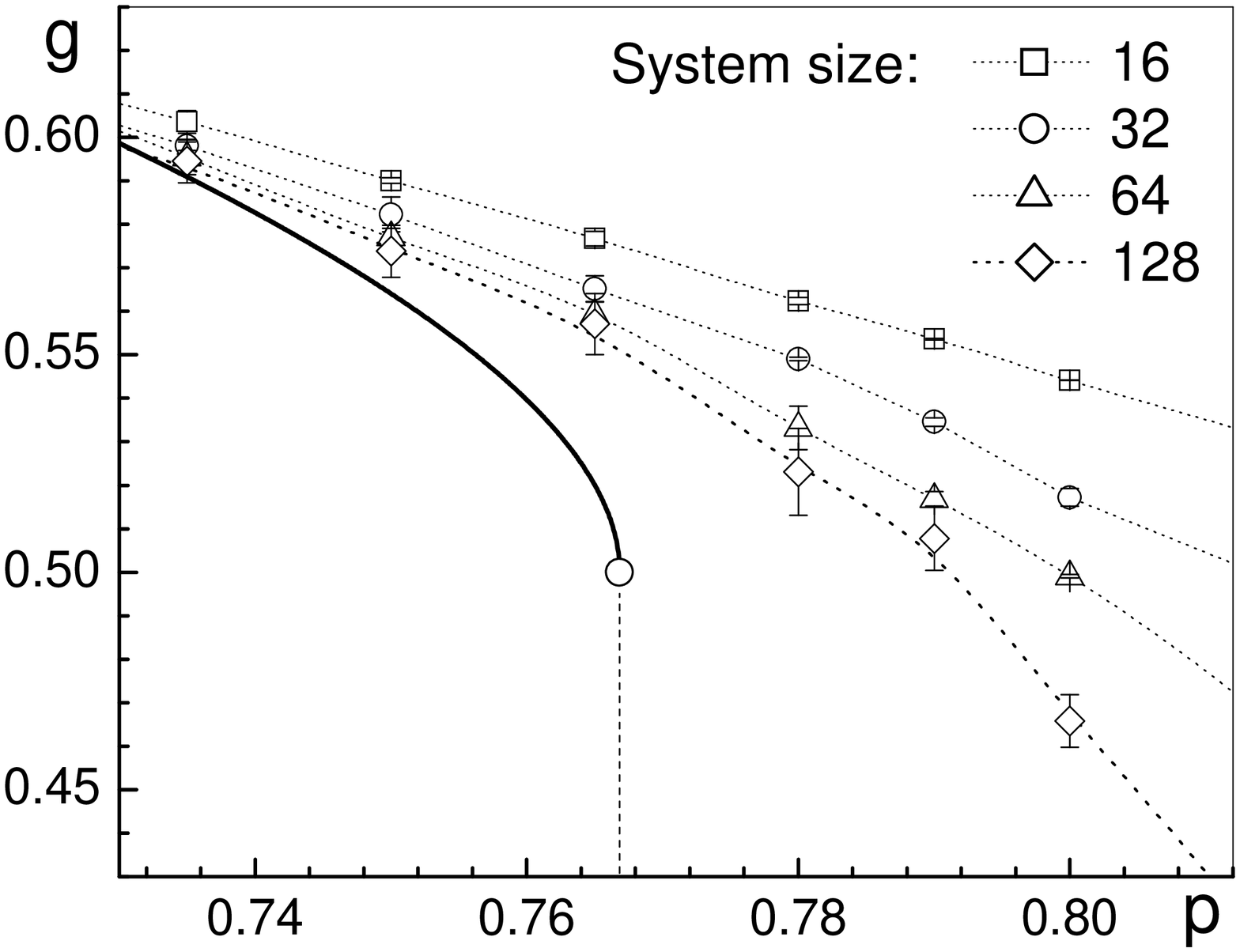}
\caption{Luttinger liquid parameter $g$ as a function of $p$.
Dotted lines are to guide the eye. The solid line is the KT
extrapolation to the infinite system size. } \label{fig:KT_g}
\end{figure}

\begin{figure}[htb]
\includegraphics[width = 0.85\columnwidth,keepaspectratio=true]{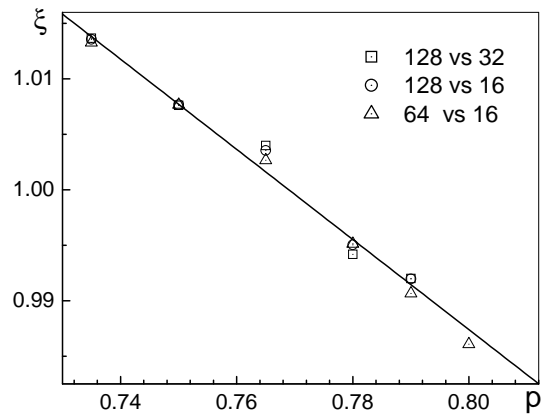}
\caption{Parameter $\xi$ for different data sets as a function of
$p$. The solid line is a linear fit. The errors are of the order
of the symbol size.} \label{fig:KT_xi}
\end{figure}

As is always the case with SF--I transitions in 1D systems, a
brute-force numeric observation of the critical $g_c$ is
problematic in view of the exponentially divergent correlation
length. We thus need to perform the finite-size analysis of the
data using Kosterlitz-Thouless renormalization-group flow:
\begin{equation}
 \int^{g(L_2)/g_c}_{g(L_1)/g_c}\, \frac{dt}{t^2 (\ln t - \xi) +
t} \, =\,4 \ln (L_1/L_2)\; . \label{KT_flow}
\end{equation}
Here $g(L)$ is the Luttinger parameter $g$ as a function of the
system size, $\xi$ is an $L$-independent microscopic parameter
(which is an analytic function of $p$). At a given $p$, the value
of $\xi$ is obtained with Eq.~(\ref{KT_flow}) from numeric values
of $g(L_1)$ and $g(L_2)$. The consistency with the
Kosterlitz-Thouless renormalization-group flow is checked by the
data collapse for different pairs of system sizes and also by the
shape of the curve $\xi(p)$ which should look as a straight line
in the vicinity of the critical point, in contrast to the $g(L,p)$
curves---as functions of $p$, at large enough $L$. These curves
should demonstrate a considerable curvature consistent with a
(slow) evolution, as $L \to \infty$, towards the jump at $p_c$
from $g=1/2$ to $g=0$. These features are seen in
Figs.~\ref{fig:KT_g}--\ref{fig:KT_xi}. After extracting the
function $\xi(p)$, the macroscopic $g(p)\equiv g(L\to \infty , p)$
limit is obtained---in accordance with Eq.~(\ref{KT_flow})---from
$~(g/g_c)\,(\ln g/g_c - \xi) = -1$.

The critical value $g=1/2$ implies a rough interface, since
roughening is apparently the only mechanism of eliminating the
effect of broken translation symmetry in the bulk. There is also a
strong argument in favor of {\it simultaneous} appearance of
superfluidity and roughening in the 1D interface. There is little
doubt that deep in the insulating phase the interface becomes
smooth ($T=0$), since roughening costs finite potential energy
that dominates over the kinetic energy in this limit. In our
simulations, we see this as the effect of ``freezing" of the
interface position at large enough $p$. [By its nature, a rough
interface experiences local fluctuations that gradually lead to
its global drift.] The zero-point roughening fluctuations in the
smooth phase are due to the specific solitons illustrated in
Fig.~\ref{fig:spinons}. These solitons shift the position of the
interface in the $x$ direction by one step. It is also seen that
they carry a topological charge associated with shifting by one
lattice period the checkerboard density wave along the interface
($y$ direction). These quasiparticles also carry a particle charge
$\pm 1/2$, as it follows, e.g., from the fact that a
single-particle hopping event translates the soliton by two
lattice spacings, see Fig.~\ref{fig:spinons}. In view of their
fractional particle charge it is conventional to call these
solitons spinons \cite{spinon}. It turns out that in the
insulating phase spinons are the {\it lowest} particle-charge
carrying elementary excitations.---The snapshots of the wordline
configuration cross-sections in planes perpendicular to the $\tau$
direction reveal structures identical to those of
Fig.~\ref{fig:spinons}. It is reasonable to assume then that the
transition from insulating to superfluid phase is due to the
proliferation of spinons, in analogy to the superfluid transition
in a standard 1D checkerboard solid, also driven by spinons.
Hence, we arrive at a picture where both superfluid and roughening
transitions occur simultaneously being driven by proliferation
(condensation) of one and the same quasiparticle mode.

In conclusion, we have demonstrated that an interface layer in a
normal solid may exhibit superfluidity in a wide range of
parameters. This result may be of direct relevance to NCMI of
solid $^4$He discovered by Kim and Chan \cite{KimChan04},
supporting the interpretation in terms of the superfluidity of
micro-crystallite interfaces. We have studied numerically
superfluid--insulator quantum phase transitions in particle-hole
symmetric interfaces in 2D and 3D models of the lattice
checkerboard solid. In 3D, the transition is in the $U(1)$
universality class implying that the interface is smooth. In 2D,
where the interface is a 1D Luttinger liquid, we observe a
Kosterlitz-Thouless type transition at Luttinger-liquid parameter
$g=1/2$, which implies that the interface is rough on the
superfluid side. We argue that the 1D interface becomes smooth
simultaneously with becoming insulating, since the onset of
superfluidity and roughening are due to proliferation of the same
quasiparticles which (i) have particle charge $\pm 1/2$, (ii)
represent defects in the checkerboard order, and (iii) are kinks
shifting the interface in the perpendicular direction by one
lattice spacing.

We are grateful to Subir Sachdev for a discussion of the results.
The research was supported by the National Science Foundation
under Grants No. PHY-0426881 and No. PHY-0426814, by NASA under
Grant NAG32870, and by PSC CUNY Grant No. 665560035.



\begin{thebibliography}{99}
\bibitem{rem1} If the interfaces in $^4$He polycrystal prove superfluid, the scenario for their solidification may
be different because of the absence of the commensurability. It is
likely that in this case the transition will generically be in the
universality class of a disordered system, with the Bose glass
type of the insulating phase.
\end{thebibliography}
\end{document}